\documentstyle[12pt,epsfig]{article}
\def\pa{\parallel}
\def\pe{\bot}

\def\bea{\begin{eqnarray}}
\def\eea{\end{eqnarray}}
\def\be{\begin{equation}}
\def\ee{\end{equation}}
 \let\b=\beta   \let\d=\delta  \let\e=\varepsilon
        
\let\m=\mu      \let\o=\omega    
   \let\t=\tau   
  \let\z=\zeta
\let\D=\Delta   \let\G=\Gamma

\title{\bf Renormalized Field Theory of Infinitely 
Driven Lattice Gases}

\author{F. de los Santos $^1$ and Miguel A. Mu\~noz $^{1,2}$
 \\ $^1$
 Institute {\em Carlos I} for Theoretical and Computational Physics\\ 
and Departamento de Electromagnetismo y F\'\i sica de la Materia\\
 18071 Granada, Spain.\\
 $^{2}$ The Abdus Salam International
 Centre for Theoretical Physics\\
P.O. Box 586, 34100 Trieste, Italy}
\begin{document}
\maketitle
\begin{abstract}
We use field theoretic renormalization group methods to study
the critical behavior of a recently proposed Langevin equation
for driven lattice gases under infinitely fast drive.
 We perform  an expansion around the upper critical
dimension, $d_c=4$, and obtain the critical exponents 
to one loop order. The main features of the 
two loop calculation are also outlined.  
The renormalized theory is shown to exhibit a behavior different
from the standard field theory for the DLG with finite driving,
 {\em i.e.\/}
it is not mean field like.
\end{abstract}

Since it was first introduced by Katz {\em et al.\/} \cite{katz},
the {\em driven lattice gas\/} model (DLG hereafter) has attracted 
considerable interest \cite{zia,marro}. 
Being one of the simplest archetypes
of non-equilibrium model its study may contribute to
pave the way for an understanding of out of equilibrium systems.
The DLG consists of a periodic regular lattice on which 
nearest-neighbor particle-hole exchanges are performed. The hopping rate
is determined by the energetics
 of the Ising Hamiltonian $H$, the coupling to
a thermal bath at temperature $T$, 
and an external uniform driving field
$\bf E$ pointing along a specific lattice axis. 
In particular,
the hoping rate depends on 
$[(\D H+ \ell E)/T]$, where $\D H$ is the energy variation 
which would be caused by 
the tried configuration change, $E=|{\bf E}|$, and
$\ell=1$ (-1) for jumps along (against) $\bf E$ and 0
otherwise (see \cite{zia,marro} for a detailed description).
The DLG exhibits a 
continuous phase transition 
from a disordered state at high $T$ to a stripe like ordered state at 
sufficiently low $T$ \cite{zia,marro}.  The nature and properties of this
transition have been largely studied in recent years. 
A new general Langevin equation has been proposed,
designed to capture the physics of the DLG at
 the critical point \cite{pre}.
 This Langevin equation
has different relevant terms
for the cases $0<E<\infty$ and
 $E=\infty$ (in which particles cannot jump against the field) 
respectively \cite{pre,jsp}. 
That is, the point $E=\infty$ behaves as
a sort of {\it tricritical point} in the parameter space where
some terms are exactly zero and the relevance
of the different operators has to be re-evaluated.
For finite values of the driving field $E$ the 
Langevin equation previously proposed by 
Janssen and Schmittmann \cite{janssen} is recovered.
It is the purpose of this paper to investigate the critical behavior 
of the DLG for $E=\infty$ in order to determine
explicitly whether the differences with the $0<E<\infty$ case are
of a relevant nature and whether, therefore, the critical behavior
is changed.

The new
Langevin equation reads \cite{pre,jsp}
\bea
\label{new}
\partial_t \phi ={e_0 \over 2}
\Big[ -\D_\pa \D_\pe \phi-\D_\pe^2 \phi+
\t \D_\pe \phi+ {g \over 3!} \D_\pe \phi^3 \Big] \nonumber \\
+\sqrt{e_0} \ \nabla_\pe \cdot \mbox{\boldmath $\xi$}_\pe
+\sqrt{e_0 \over 2} \ \nabla_\pa \xi_\pa,
\eea
where $\nabla_\pa $ ($\nabla_\pe$) is 
 the gradient operator in 
the direction parallel (perpendicular) to the electric field, and
the noise satisfies
\bea
\langle \mbox{\boldmath $\xi$}({\bf x},t) \rangle &=&0, \nonumber \\
\langle \mbox{\boldmath $\nabla$}
\cdot \mbox{\boldmath $\xi$}({\bf x},t)
\cdot \mbox{\boldmath $\nabla$}' \cdot
\mbox{\boldmath $\xi$}({\bf x}',t') \rangle
&=&-\mbox{\boldmath $\nabla$}^2 \d({\bf x}-{\bf x}') \d(t-t').
\eea
This equation is analogous to a model B in the direction perpendicular
to the field (where the energy takes into account the interaction with the
parallel direction through the crossed derivatives term),
coupled to a simple random diffusion mechanism in the
parallel direction.  This means that {\it the relevant ingredient 
of the infinite driving is the anisotropy it introduces,
 while the directionality of the
flux is irrelevant from a renormalization group point of view}. 
A very similar equation has been proposed to
describe the {\it Freedericksz transition} in nematic liquids, 
and general asymmetric two-dimensional pattern formation \cite{nematic}.

In order to renormalize this equation,
 following standard methods \cite{bausch},
let us introduce a Martin-Siggia-Rose
response field $\tilde \phi$ and recast 
Eq. (\ref{new}) as a dynamical functional \cite{GF}, the
associated action of which is
\bea
{\cal L}(\tilde \phi, \phi)= \int d^dx dt \Bigg\{
\tilde \phi \Big[ \partial_t -{e_0 \over 2} (-\D_\pa \D_\pe
-\D_\pe^2 +\t \D_\pe) \Big] \phi
-{e_0 \over 2} {g \over 3!} \tilde \phi \D_\pe \phi^3 \nonumber \\
-{e_0 \over 2} \tilde \phi \Big (\nabla_\pe^2 + {1 \over 2}
\nabla_\pa^2 \Big)
\tilde \phi \Bigg\}.
\label{funcional}
\eea

\begin{figure}
%

\centerline{
\epsffile{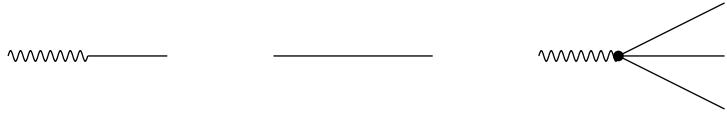}}
\caption{Elements of perturbation theory: the response and correlation 
propagators and the four-point vertex. $\tilde \phi$-legs are indicated by 
a wiggle.}
\end{figure}
 
The free propagators are 
\bea
G_{02}^0({\bf k},\o) &=& {-e_0 ( k_\pe^2+{1 \over 2}k_\pa^2) \over
\o^2+({e_0 \over 2})^2 k_\pe^4(k^2 + \t)^2},
\nonumber \\
G_{11}^0({\bf k},\o) &=& {1 \over i\o+{e_0 \over 2} k_\pe^2(k^2 + \t)},
\eea
and the vertex is: 
$- e_0 g /12 \ k_\pe^2$. 
These elements can be represented diagrammatically as in 
Figure 1 (wavy legs symbolize response fields; straight 
lines stand for density fields).

In order to renormalize the theory, one has to look for the primitive
divergences in a perturbation expansion.
If $\G_{\tilde n n}$ denotes a one-particle irreducible vertex
function with $\tilde n$ external $\tilde \phi$-legs and $n$
external $\phi$-legs, only $\G_{11}$ and $\G_{13}$ are found
to possess primitive divergences. The Feynman diagrams contributing
to these vertex functions (shown in Figure 2) 
are topologically identical to model B 
graphs \cite{bausch}.
 However, the bare correlation and response propagators that
follow from  Eq. (\ref{funcional}) are anisotropic, in contrast to their
counterparts in model B \cite{bausch}. 

\begin{figure}[t]

\centerline{
\epsfbox{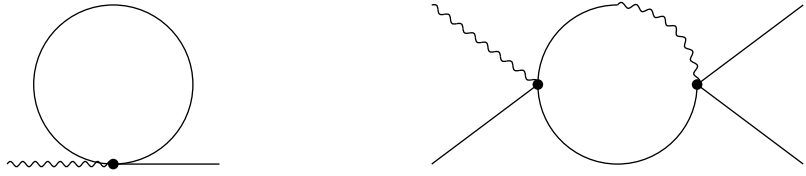}}
\caption{One loop diagrams contributing to
$\G_{11}$ (left) and $\G_{13}$ (right).}
\label{diagdiv}
\end{figure}

To one loop in $\e=4-d$, the ultraviolet divergences in $\G_{11}$ and 
$\G_{13}$ lead to the renormalization of $\t$ and $u$, the latter
being the dimensionless coupling constant $u \equiv A_{\e} \t^{-\e/2} g$.
$A_\e$ is a numerical factor to be defined below.
We define renormalized parameters $\t_R$ and $u_R$ by
$\t_R = Z_\t \t$ and 
$u_R = Z_u u $.
Given that the leftmost diagram in Figure 2 does not depend
on external moments nor frequencies, therefore
the derivatives of $\G_{11}$ with respect
to them vanish, and no extra (field) renormalizations are 
required.
The $Z$ factors are determined by the following normalization conditions
\bea
\partial_{k_\bot^2} \G_{11}^R \vert_{NP} &=& {e_0 \over 2} \t_R, \nonumber \\
\partial_{k_\bot^2} \G_{13}^R \vert_{NP} &=& {e_0 \over 2} \t^{\e/2} A_\e^{-1}
u_R.
\eea
A convenient choice for the normalization point NP is
${k}_i= \o_i =0$ and $\t =\mu^2$, where $\mu$ is an arbitrary momentum
scale. To one loop, we find 
\bea
\G_{11} (w, {\bf k}) &=& i\o +{e_0 \over 2}k_\pe^2 (k^2+\t)+D_1, \nonumber \\
\G_{13} (w, {\bf k} ) &=& {e_0 \over 2}k_\pe^2 g + D_2,
\eea
where $D_1$($D_2$) corresponds to the algebraic expression of 
the left(right) diagram in Figure \ref{diagdiv}.
A calculation in dimensional regularisation \cite{amit,GF} yields 
\bea
D_1&=& {5 g e_0\over 128 \pi^2} k_\bot^2 {\t^{1-\e/2} 
\over \e}, \nonumber \\
D_2&=& 
-{5 g^2 e_0\over 128 \pi^2} k_\bot^2 {\t^{-\e/2}  \over \e}.
\eea
After setting $A_\e = 5/64 \pi^2$, one obtains
\bea
\partial_{k_\bot^2} \G_{11} \vert_{NP} &=&
{e_0 \over 2} \t +  {5 g e_0\over 128 \pi^2} {\t^{1-\e/2}
 \over \e}= {e_0 \over 2} \t \bigg[1+ {5 \t^{-\e/2}
\over 64 \pi^2} {g \over \e}
\bigg]= {e_0 \over 2} \t \bigg[ 1+{u \over \e} \bigg], \nonumber \\
\partial_{k_\bot^2} \G_{13} \vert_{NP} &=&
{e_0 \over 2} A_\e^{-1} \t^{\e/2} u \bigg[1- {u \over \e}\bigg],
\eea
which entails
\bea
Z_\t= 1+{u \over \e} + O(u^2), \nonumber \\
Z_u= 1-{u \over \e}+ O(u^2).
\eea
The renormalization group equation obtained
after requiring invariance of the bare irreducible vertex functions
upon changes on the normalization point reads
\be
\Big[ \m \partial_\m +\b \partial_{u_R} + \z \partial_{\t_R}
\Big] \G_{\tilde n n}^R =0,
\ee
where the renormalization group functions are defined in the usual way:
$\b(u_R) \equiv \m \partial_\m u_R$, 
and $\z(u_R) \equiv \m \partial_\m (\ln \t_R)$.
A straightforward calculation then leads to 
\bea
\b (u_R) &=&-\e u_R + u_R^2 + O(u_R^3), \nonumber \\
\z (u_R) &=& 2-u_R +O(u_R^2),
\eea
from which one can determine the location and stability of the fixed
points. 
To this order, apart from the trivial mean-field result $u_R^*=0$, 
a nontrivial, infrared stable, fixed point $u^*_R=\e$ emerges.
 This fixed point controls the
critical behavior of the theory below four dimensions. 

\begin{figure}
\centerline{
\epsfbox{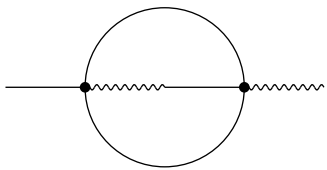}}
\label{dosloops}
\caption{Two loop contribution to $\G_{11}$.}
\end{figure}

Now we proceed with the calculation of the associated critical exponents.
We first note that, as indicated above,
 no renormalization of the fields 
$\tilde \phi$, $\phi$ has been required. Therefore,
in particular, the anomalous dimension of $\phi$ vanishes up to 
one loop, {\em i.e.\/}, 
$\eta =0+ O(\e^2)$.
Concerning the exponent $\nu_{\perp}$, which controls the divergence of
the correlation length with temperature \cite{noten},
 we simply have 
$\nu_{\perp}= \zeta(u_R^*)^{-1}$, and $\nu_{\perp}=1/2+\e /4 + O(\e^2)$. 
This is to be compared with $\nu =1/2$, the value obtained by Janssen  and 
Schmittmann in the standard field theory \cite{janssen}. 
{\it This result demonstrates that
the continuous version Eq. (\ref{new}) of the DLG with
 $E=\infty$ is not mean-field like but characterizes
a universality class other than the 
one in \cite{janssen}}.
Since there are no dangerous
 irrelevant operators in Eq. (\ref{funcional}) 
 standard scaling laws apply (contrary to the case in \cite{janssen}).
Therefore the exponents are related to each other and
estimating $\eta$ and $\nu_{\perp}$ is sufficient to 
deduce all the other exponents.
 For instance, the order parameter exponent $\b$
can be written as
$\b ={\nu_{\perp} \over 2} (d-2+\eta)$ \cite{amit}, 
and we have $\b =1/ 2 +O(\e^2)$.

The previous results concern the one loop approximation.
The two loop calculation  
presents an interesting
new feature, namely, that the scaling becomes fully anisotropic.
In fact,
Figure 3 reveals that, contrary to what happens
to one loop order, the two loops correction 
to $\G_{11}$ depends on external frequencies and momenta; and 
in absence  of any symmetry between parallel and perpendicular derivatives,
one can easily convince himself by simple inspection that
$\partial_{k_{\perp}^4} \G_{11} \neq
 \partial_{k_{\perp}^2}  
 \partial_{k_{\pa}^2} \G_{11} $. In order to absorb these two
different divergences, one is constrained to renormalize
 the parallel and the perpendicular
 momenta in a different way: if $k_{\perp} \rightarrow 
l k_{\perp}$ then $k_{\pa} \rightarrow 
l^{1+\gamma} k_{\pa}$ (where $\gamma \propto \e^2$
 can be determined by explicitly computing 
the derivatives of the diagram in Figure 3).
 The scaling law has to 
be rewritten as $\beta={\nu_{\perp}/2 (d-2+\gamma +\eta})$, and all 
the exponents become non mean-field in this approximation.

Summing up, we have performed the renormalization of the 
field theory in \cite{pre} for the DLG under 
an infinitely large driving field. The renormalization procedure
yields results essentially different from those for a 
finite field. In particular, corrections to mean field
are observed explicitly in the one loop approximation for 
the exponent $\nu_{\perp}$. 
Anisotropic exponents and a non-mean-field exponent 
$\beta$ appear from simple
arguments based on the analysis up to two loop diagrams. These 
severe differences with respect to the finite driving field case  
call for extensive computational 
simulations to observe numerically the physical differences
between both cases. 

{\bf Acknowledgements} It is a pleasure to acknowledge J. Marro,
J. L. Lebowitz and P.L. Garrido
 for useful discussions, and E. Hern\'andez-Garc{\'\i}a 
for pointing out reference \cite{nematic} to us.
 This
work has been partially supported by the 
European Network Contract ERBFMRXCT980183
 and by the Ministerio de Educaci\'on
under project DGESEIC, PB97-0842.

\end{document}